\documentclass[sigconf, screen]{acmart}
\usepackage{subcaption}
\usepackage{booktabs}
\usepackage{multirow}
\AtBeginDocument{%
  }

\setcopyright{acmlicensed}
\copyrightyear{2018}
\acmYear{2018}
\acmDOI{XXXXXXX.XXXXXXX}
\acmConference[Conference acronym 'XX]{Make sure to enter the correct
  conference title from your rights confirmation email}{June 03--05,
  2018}{Woodstock, NY}
\acmISBN{978-1-4503-XXXX-X/2018/06}
\setcopyright{none}
\renewcommand\footnotetextcopyrightpermission[1]{}
\begin{document}

\title{Contrastive Learning for Continuous Touch-Based Authentication}


\author{Mengyu Qiao}
\authornote{Corresponding author.}
\affiliation{%
  \institution{ North China University of Technology}
  \city{Beijing}
  \country{China}
}
\email{myuqiao@ncut.edu.cn}

\author{Yunpeng Zhai}
\affiliation{%
  \institution{North China University of Technology}
  \city{Beijing}
  \country{China}}
\email{2023322030149@mail.ncut.edu.cn}

\author{Yang Wang}
\affiliation{%
  \institution{Ultramain Systems, Inc.}
  \city{Albuquerque}
  \state{NM}
  \country{USA}}
  \email{ywang@ultramain.com}

\renewcommand{\shortauthors}{Qiao et al.}

\begin{abstract}
  Smart mobile devices have become indispensable in modern daily life, where sensitive information is frequently processed, stored, and transmitted—posing critical demands for robust security controls. Given that touchscreens are the primary medium for human-device interaction, continuous user authentication based on touch behavior presents a natural and seamless security solution. While existing methods predominantly adopt binary classification under single-modal learning settings, we propose a unified contrastive learning framework for continuous authentication in a non-disruptive manner. Specifically, the proposed method leverages a Temporal Masked Autoencoder to extract temporal patterns from raw multi-sensor data streams, capturing continuous motion and gesture dynamics. The pre-trained TMAE is subsequently integrated into a Siamese Temporal-Attentive Convolutional Network within a contrastive learning paradigm to model both sequential and cross-modal patterns. To further enhance performance, we incorporate multi-head attention and channel attention mechanisms to capture long-range dependencies and optimize inter-channel feature integration. Extensive experiments on public benchmarks and a self-collected dataset demonstrate that our approach outperforms state-of-the-art methods, offering a reliable and effective solution for user authentication on mobile devices.
\end{abstract}



\received{20 February 2007}
\received[revised]{12 March 2009}
\received[accepted]{5 June 2009}

\maketitle

\section{Introduction}

The widespread adoption of smartphones has profoundly transformed human–device interaction in both personal and professional spheres. These devices now handle not only communication and entertainment but also sensitive tasks such as financial transactions and private data storage. As dependence on mobile technology continues to grow, safeguarding user data has become an increasingly critical priority.

Conventional authentication methods—such as PIN codes, passwords, and static biometric scans—are widely used for convenience. However, they remain susceptible to various attacks, including credential theft and biometric spoofing \cite{silasai2020study}. To overcome these vulnerabilities, behavioral biometrics have emerged as a promising supplementary layer for continuous authentication \cite{lamb2020swipe}. Unlike traditional methods, behavioral biometrics continuously monitor users’ interactions with their devices, enabling persistent and dynamic identity verification.

Among the various behavioral biometric modalities, touch dynamics has garnered increasing scholarly attention owing to its unobtrusiveness and temporal consistency. Unlike gait or motion-based systems, which rely on sensors that are susceptible to changes in user posture or contextual conditions, touch-based authentication leverages interaction-specific features such as tap pressure, swipe velocity, and temporal rhythm \cite{shen2015performance, shen2022mmauth}. These characteristics exhibit relatively low sensitivity to environmental and physiological variability, thereby rendering touch dynamics a viable candidate for robust, real-world continuous authentication systems.

Despite progress in behavioral biometrics, many existing methods rely on handcrafted features and traditional classifiers like SVMs, without fully exploiting the temporal structure inherent in user interactions. This limits their ability to capture sequential dependencies critical for modeling user-specific behavior over time. Ignoring such temporal dynamics reduces the discriminative power and robustness of authentication systems \cite{shen2017performance}.

To address these challenges, we propose TouchSeqNet—a Siamese time-series framework that integrates a pre-trained Temporal-Atten\\tive Convolutional Network  (TACN) with contrastive learning. The proposed architecture employs a self-supervised pretraining phase based on a Temporal Masked Autoencoder (TMAE), which reconstructs masked segments of time-series data to learn generalizable temporal representations. These representations are subsequently transferred to a Siamese network, where TACN refines the features through dilated causal convolutions and multi-head self-attention mechanisms. Additionally, a finger-channel attention module adaptively highlights the most discriminative features across input sequences. Finally, the representations of sample pairs are concatenated and passed through a classification head to determine identity similarity.

To evaluate the model, we introduce Ffinger, a new dataset comprising touch dynamics from 29 users. Ffinger captures diverse interaction patterns across users. Additionally, we benchmark performance using two widely adopted datasets—BioIdent and Touchalytics—ensuring fair comparison under standard protocols. It shows that our method is superior to existing gesture detectors and time series classification baselines and achieves stateof-the-art performance. In summary, the contributions of this paper are threefold as below:
\begin{itemize}
    \item We reformulate continuous authentication as a contrastive learning task by employing a Siamese network architecture, offering a novel perspective in this field.
    \item We propose a Temporal Masked Autoencoder for self-superv\\ised pertaining, which effectively captures fine-grained temporal patterns associated with continuous motion and gesture dynamics, enabling the extraction of robust and generalizable representations.
    \item We propose a Temporal-Attentive Convolutional Network, which incorporates dilated causal convolutions, multi-head self-attention, and channel attention mechanisms to further enhance the network's ability to capture long-range temporal dependencies and optimize feature integration.
\end{itemize}

\begin{figure*}[t]
	\centering
	\includegraphics[width=0.9\textwidth]{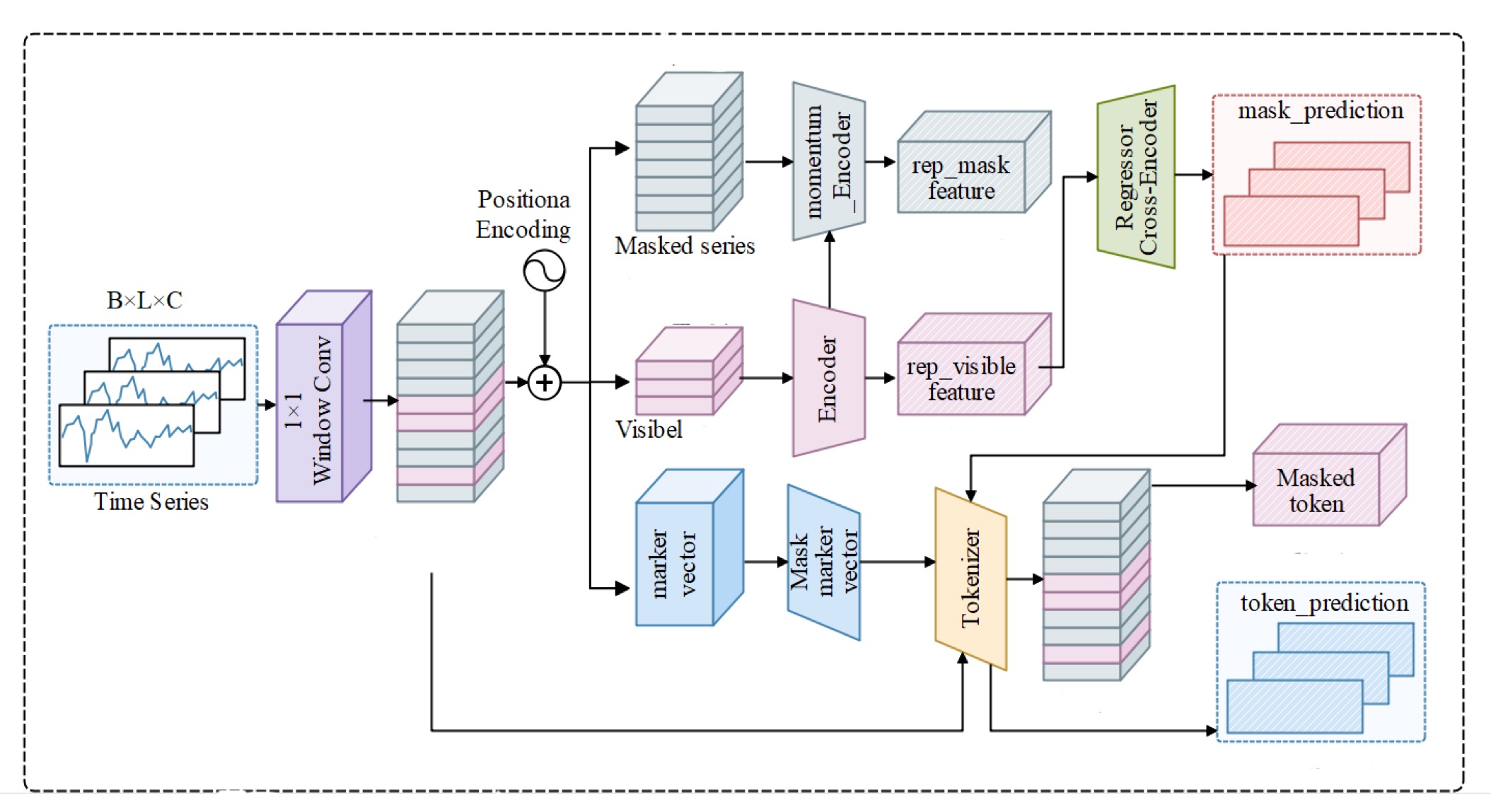}
	\caption{TMAE Model Architecture}
        \Description{An illustration of the pre-training process, where a time series input is masked and reconstructed using a transformer-based encoder-decoder architecture.}
	\label{fig:Pre-training}
\end{figure*}

\section{Related Work}

A wide range of behavioral biometric modalities have been explored for continuous user authentication on mobile devices, including touch dynamics~\cite{fierrez2018benchmarking, frank2012touchalytics, tolosana2020biotouchpass2, wang2022using, song2022integrating}, motion sensor signals~\cite{centeno2018mobile, shen2017performance, amini2018deepauth, mekruksavanich2021deep, neverova2016learning}, keystroke dynamics~\cite{stragapede2022mobile, acien2021typenet, sun2019smartphone, stragapede2023mobile}, and gait patterns~\cite{wang2021framework, middya2021privacy}. These studies have demonstrated the effectiveness of mobile behavioral authentication in both constrained and unconstrained environments.

Among these modalities, touch-based biometrics have drawn significant attention due to their unobtrusiveness and stability. Frank et al.~\cite{frank2012touchalytics} pioneered large-scale evaluation of touch features, achieving 2\% EER with multi-gesture fusion. Tolosana et al.~\cite{tolosana2020biotouchpass2} introduced MobileTouchDB and proposed a Siamese LSTM-DTW framework. More recent works~\cite{wang2022using, song2022integrating} leveraged CNNs and multimodal representations to improve classification accuracy.

Motion and sensor-based methods also show promising performance, especially when combining CNNs or LSTMs with accelerometer and gyroscope data~\cite{centeno2018mobile, amini2018deepauth}. DeepConvLSTM~\cite{mekruksavanich2021deep} and clock-variant RNNs~\cite{neverova2016learning} have been applied to capture temporal dependencies in such signals. Similarly, keystroke dynamics have been modeled with RNNs~\cite{acien2021typenet, sun2019smartphone} and Transformer hybrids~\cite{stragapede2023mobile}, while gait recognition systems have adopted metric learning~\cite{wang2021framework} and deep CNNs~\cite{middya2021privacy} for robust identification.

Despite encouraging progress, many existing methods rely on static or handcrafted features and shallow models, limiting their ability to capture the sequential and dynamic nature of user behavior. Moreover, few approaches incorporate self-supervised learning or pretraining strategies, which are crucial for generalization in real-world deployment.

\section{Data Acquisition and Processing}

\subsection{Ffinger Dataset}

\textit{Ffinger} containing interaction data from 29 participants. Each participant performed both predefined gestures—seven structured multi-touch tasks denoted by \( \{a, b, c, d, e, f, g\} \)—and free-form gestures, which allowed users to draw arbitrary patterns. For each sample, trajectories from all five fingers were simultaneously recorded.

Each finger's trajectory is represented as a 7-channel time series, including the following features:

\begin{itemize}
    \item \( x_i, y_i \): spatial coordinates of the touch point at time \( t_i \),
    \item \( t_i \): timestamp,
    \item \( p_i \): applied pressure,
    \item \( s_i \): touch area size,
    \item \( v_i \): instantaneous velocity,
    \item \( d_i \): movement direction,
\end{itemize}

This design captures both structured and natural usage scenarios, enabling fine-grained modeling of user behavior.

\subsection{Data Processing}

To ensure consistency across datasets, we selected five core features as model input: timestamp (\( T_i \)), horizontal and vertical positions (\( X_i, Y_i \)), applied pressure (\( P_i \)), and contact area (\( A_i \)). Each time step is represented as:
\begin{equation}
\mathbf{X}_i = \left[ T_i, X_i, Y_i, P_i, A_i \right]
\end{equation}

\paragraph{First-Order Differencing.}
To emphasize motion dynamics and reduce temporal redundancy, we compute the differences in time and position:
\begin{equation}
T_i' = T_i - T_{i-1}, \quad X_i' = X_i - X_{i-1}, \quad Y_i' = Y_i - Y_{i-1}
\end{equation}
with \( T_1 = X_1 = Y_1 = 0 \) by default.

\paragraph{Z-Score Normalization.}
To mitigate scale disparities across feature dimensions, we apply Z-score normalization to pressure and contact area features within each gesture sample:
\begin{equation}
P'_i = \frac{P_i - \mu_P}{\sigma_P}, \quad A'_i = \frac{A_i - \mu_A}{\sigma_A}
\end{equation}
where \( \mu \) and \( \sigma \) are the mean and standard deviation computed within the current sample. This improves training stability while preserving relative intra-sample variation.

\begin{figure}[htbp]
    \centering
    \begin{subfigure}[b]{0.23\textwidth}
        \centering
        \hspace*{-0.8cm}
        \includegraphics[width=\textwidth]{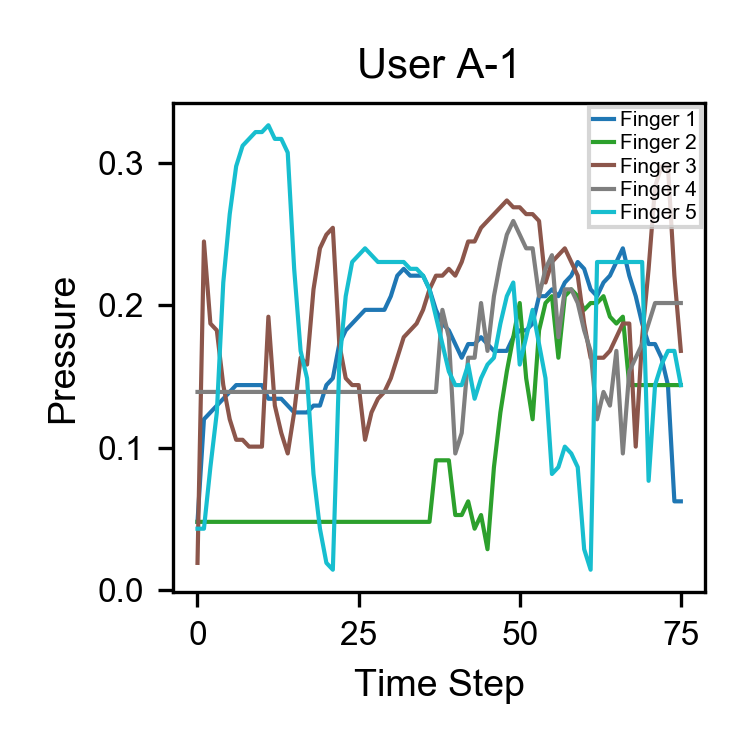}
        \caption{}
        \label{fig:pressure_user1_a}
    \end{subfigure}%
    \hfill
    \begin{subfigure}[b]{0.23\textwidth}
        \centering
        \hspace*{-0.8cm}
        \includegraphics[width=\textwidth]{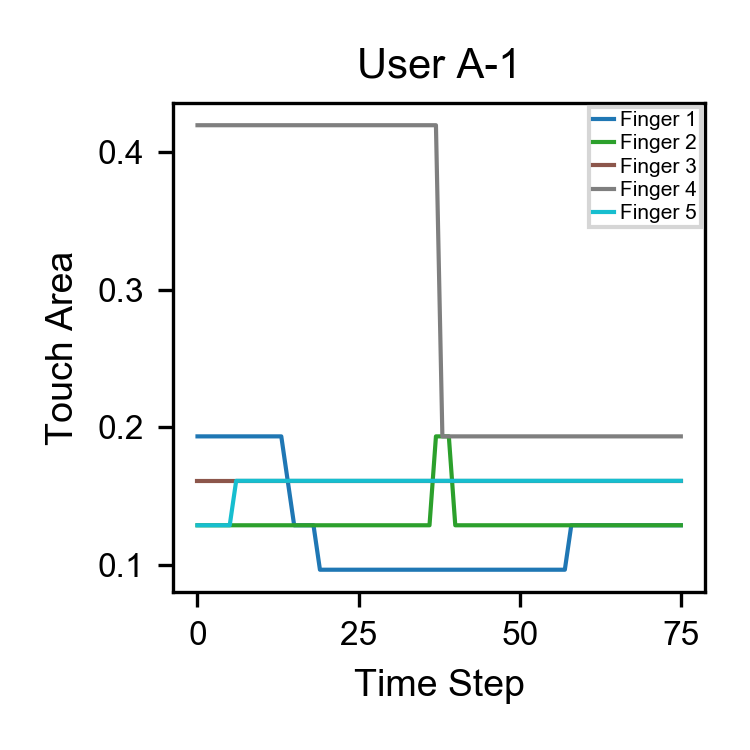}
        \caption{}
        \label{fig:area_user1_a}
    \end{subfigure} \\[-0.1cm]  

    \begin{subfigure}[b]{0.23\textwidth}
        \centering
        \hspace*{-0.8cm}
        \includegraphics[width=\textwidth]{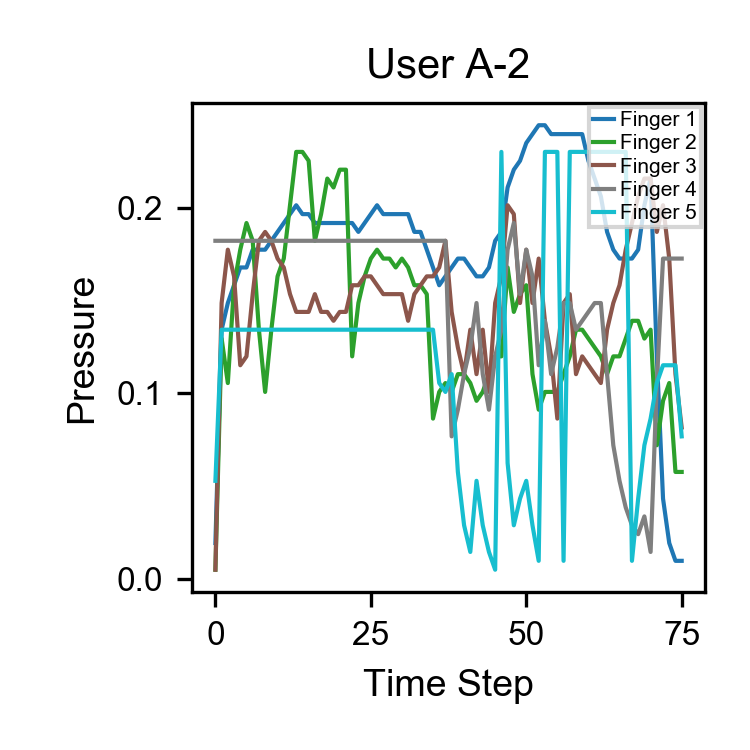}
        \caption{}
        \label{fig:pressure_user2_b}
    \end{subfigure}%
    \hfill
    \begin{subfigure}[b]{0.23\textwidth}
        \centering
        \hspace*{-0.8cm}
        \includegraphics[width=\textwidth]{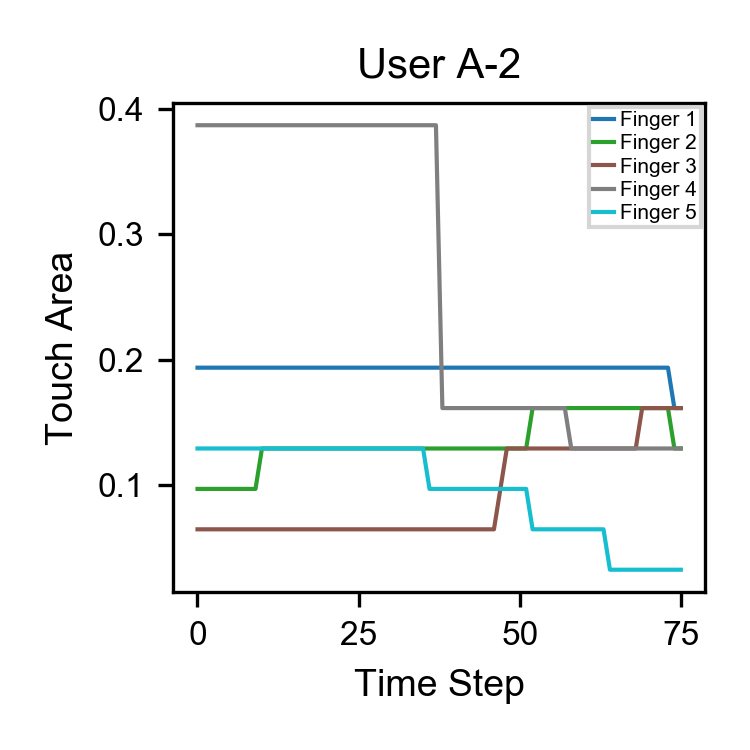}
        \caption{}
        \label{fig:area_user2_b}
    \end{subfigure} \\[-0.1cm]  

    \begin{subfigure}[b]{0.23\textwidth}
        \centering
        \hspace*{-0.8cm}
        \includegraphics[width=\textwidth]{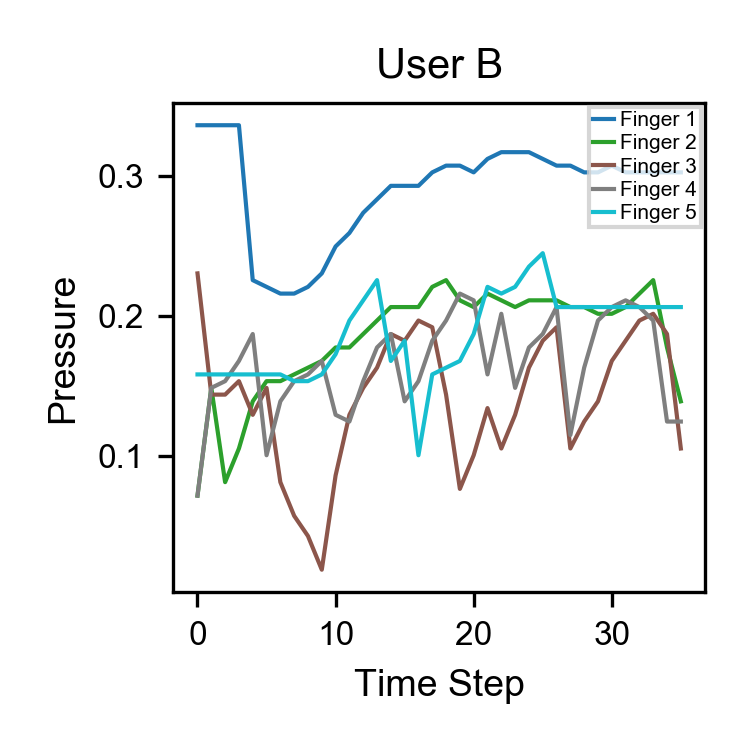}
        \caption{}
        \label{fig:pressure_user2_b_alt}
    \end{subfigure}%
    \hfill
    \begin{subfigure}[b]{0.23\textwidth}
        \centering
        \hspace*{-0.8cm}
        \includegraphics[width=\textwidth]{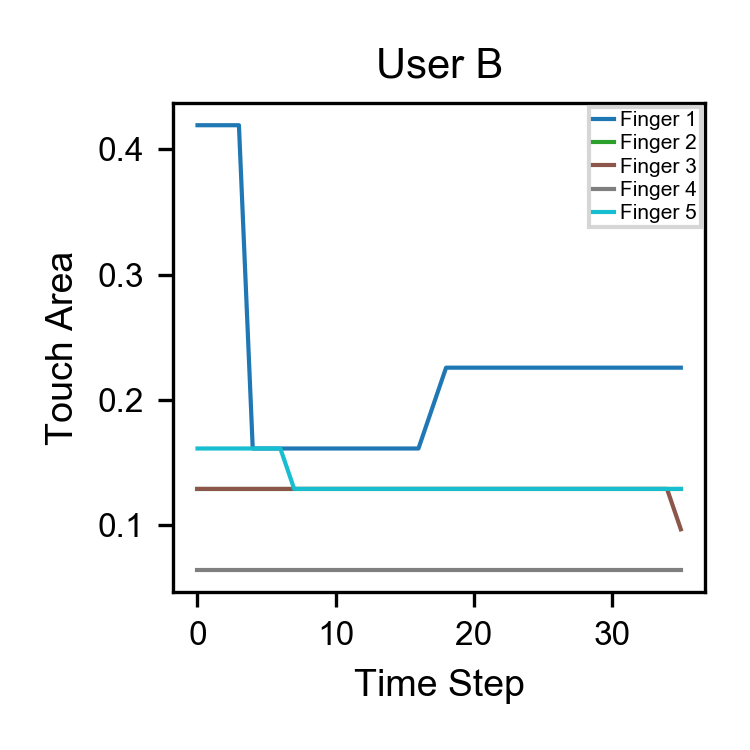}
        \caption{}
        \label{fig:area_user2_b_alt}
    \end{subfigure}

    \caption{Comparison of touch dynamics for different gestures across users. (a), (b), (c), and (d) show the pressure and touch area of two instances of the same user. (e) and (f) represent the sequences of a different user.}
    \label{fig:comparison}
\end{figure}

Figure~\ref{fig:comparison} provides a comparison of touch dynamics for gestures performed by two users, showing the high similarity between gestures of the same user and low similarity between gestures of different users.

\section{TMAE Pre-trained Model Architecture}

In this section, we present the architecture of the Temporal Masked Autoencoder (TMAE), As illustrated in Figure~\ref{fig:Pre-training}, which serves as the self-supervised pretraining backbone for time-series representation learning. The model first processes the raw time-series input through a window-based convolutional projection to extract local feature sequences. we split the feature sequences into two branches: one branch generates discrete token embeddings through a learnable tokenizer, and these embeddings are divided into visible and masked parts; the other branch directly divides the feature sequences into visible and masked parts for representation prediction. Specifically, the visible features are encoded by a Transformer encoder to capture contextual dependencies, and the masked segments are processed by a momentum-updated encoder to produce target representations. A cross-attention regressor then predicts the masked representations using information from the visible tokens, while a decoder reconstructs the discrete codewords. 

\subsection{Touch Dynamic Feature Representation}

We introduce the methods for representing touch dynamic sequences through window slicing and embedding via the Tokenizer \cite{lee2023learning}.

In order to capture meaningful temporal dependencies across different time scales. We employ a window-slicing strategy to segment the continuous touch sequence into smaller, fixed-length sub-series \cite{imani2021multi}. 

Formally, let the touch sequence be represented as \( X = \{x_1, x_2,\\ \dots, x_T \} \in \mathbb{R}^{T \times C} \), where \( T \) is the length of the sequence and \( C \) is the number of channels (features such as x and y coordinates, pressure, etc.). We slice the sequence into non-overlapping windows of size \( \sigma \), where each window \( s_{i:j} = \{ x_i, x_{i+1}, \dots, x_{i+\sigma} \} \) corresponds to a sub-series of length \( \sigma \). The number of resulting sub-series is determined by \( d = \lceil T / \sigma \rceil \), Then, the original sequence is encoded as \( Z = \{z_1, z_2, \dots, z_d \} \in \mathbb{R}^{d \times m} \), where \( d \) is the new sequence length and \( m \) is the Model embedding dimension.

This strategy reduces temporal redundancy while ensuring that each sub-series contains enough semantic information, further enhancing the self-supervised learning process.

We convert each sub-series window \( Z \) into discrete embeddings using a Tokenizer~\cite{xiao2024gaformer}, which maps raw segments into a compact latent space. Unlike handcrafted features, the Tokenizer supports end-to-end learning for effective representation of touch dynamics.

The Tokenizer module transforms each input sub-series \( s_{i:j} \in \mathbb{R}^{\sigma \times C} \) into a continuous embedding representation \( E \in \mathbb{R}^{\sigma \times m} \), where \( \sigma \) denotes the window length, \( C \) is the number of input channels, and \( m \) is the embedding dimension.

Each embedding \( E_i \in \mathbb{R}^m \) is then projected into a vocabulary space of size \( K \) using a linear layer:
\begin{equation}
\mathbf{p}_i = \text{softmax}(W E_i + b), \quad i \in [\sigma]
\end{equation}
where \( W \in \mathbb{R}^{m \times K} \) and \( b \in \mathbb{R}^K \) are learnable parameters. This projection generates a probability distribution over the codebook entries.

To enable end-to-end differentiability, we adopt the Gumbel-Softmax trick to approximate discrete sampling during training. The final discrete token \( T_i \) for each position is selected via:
\begin{equation}
T_i = \arg\max_j \mathbf{p}_i^{(j)}, \quad j \in [K]
\end{equation}

Through this process, the windowed feature sequence Z is mapped into a discrete token sequence \( T = \{T_1, T_2, \dots, T_d\} \), where each token encodes a local temporal pattern. These discrete representations provide a compact and informative abstraction of the raw input, facilitating downstream tasks such as classification and anomaly detection.

\subsection{Masking Strategies}

We describe the masking strategy used in the TMAE model as self-supervised learning \cite{devlin2019bert}. The primary objective of this strategy is to reconstruct the hidden representations of masked windows and predict their corresponding discrete tokens, thereby enabling the learning of temporally structured semantic features from touch dynamic data.

To retain the positional information of each token after windowing, we first add positional encoding to the sequence embedding \( Z \) \cite{vaswani2017attention}, resulting in:

\begin{equation}
Z^{\text{p}} = Z + \text{PositionEncoding}(Z), \quad Z^{\text{p}} \in \mathbb{R}^{d \times m}
\end{equation}

Next, we split the window-convolved sequence \( Z \) into visible and masked representations. Let \( v_{\text{index}} \) denote the indices of the visible representations, and \( m_{\text{index}} \) denote the indices of the masked representations. The corresponding parts of the sequence can be denoted as:

\begin{equation}
Z_v = Z[v_{\text{index}}] \in \mathbb{R}^{d_v \times m}, \quad Z_m = Z[m_{\text{index}}] \in \mathbb{R}^{d_m \times m}
\end{equation}

where \( d_v + d_m = d \), indicating that the overall sequence length remains unchanged.

To ensure consistency, we similarly divide the discrete token sequence \( T \) into:

\begin{equation}
T_v = T[v_{\text{index}}] \in \mathbb{R}^{d_v}, \quad T_m = T[m_{\text{index}}] \in \mathbb{R}^{d_m}
\end{equation}

During training, to represent the masked positions, we introduce a learnable mask token embedding vector \( \mathbf{m} \in \mathbb{R}^{m} \) \cite{he2021mae}. This vector is repeated for each masked position and combined with its positional encoding to form the masked input representation:

\begin{equation}
E_{\text{mask}} = \mathbf{m} \cdot \mathbf{1} + \text{PositionEncoding}(Z_m), \quad E_{\text{mask}} \in \mathbb{R}^{d_m \times m}
\end{equation}
	
The mask token serves as a placeholder for the missing information and guides the model to learn to reconstruct the masked parts \( Z_m \) based on the contextual information from the visible part \( Z_v \).

\subsection{Self-supervised Regression}

We perform two core pretext tasks: masked representation regression and discrete codeword prediction. In the following, we detail the implementation of these objectives and explain how our encoding strategy and momentum-based updates facilitate effective representation learning.

\subsubsection{Multi-head Attention}

The encoder in TMAE adopts the Multi-head Attention (MHA) mechanism~\cite{vaswani2017attention, xiao2024gaformer} to capture temporal dependencies across multiple subspaces.

Given the input sequence \( X \in \mathbb{R}^{T \times m} \), each attention head \( h \) computes:

\begin{equation}
Q_h = XW_h^Q,\quad K_h = XW_h^K,\quad V_h = XW_h^V
\end{equation}

\begin{equation}
\text{head}_h = \text{softmax}\left(\frac{Q_h K_h^\top}{\sqrt{d_{\text{head}}}}\right)V_h
\end{equation}

The outputs of all heads are concatenated and linearly transformed:

\begin{equation}
\text{MultiHead}(Q, K, V) = \text{Concat}(\text{head}_1,\dots,\text{head}_H)W^O
\end{equation}

This structure allows the encoder to model complex temporal patterns in user interactions by leveraging multiple perspectives in parallel.

\begin{figure*}[t]
	\centering
	\includegraphics[width=0.9\textwidth]{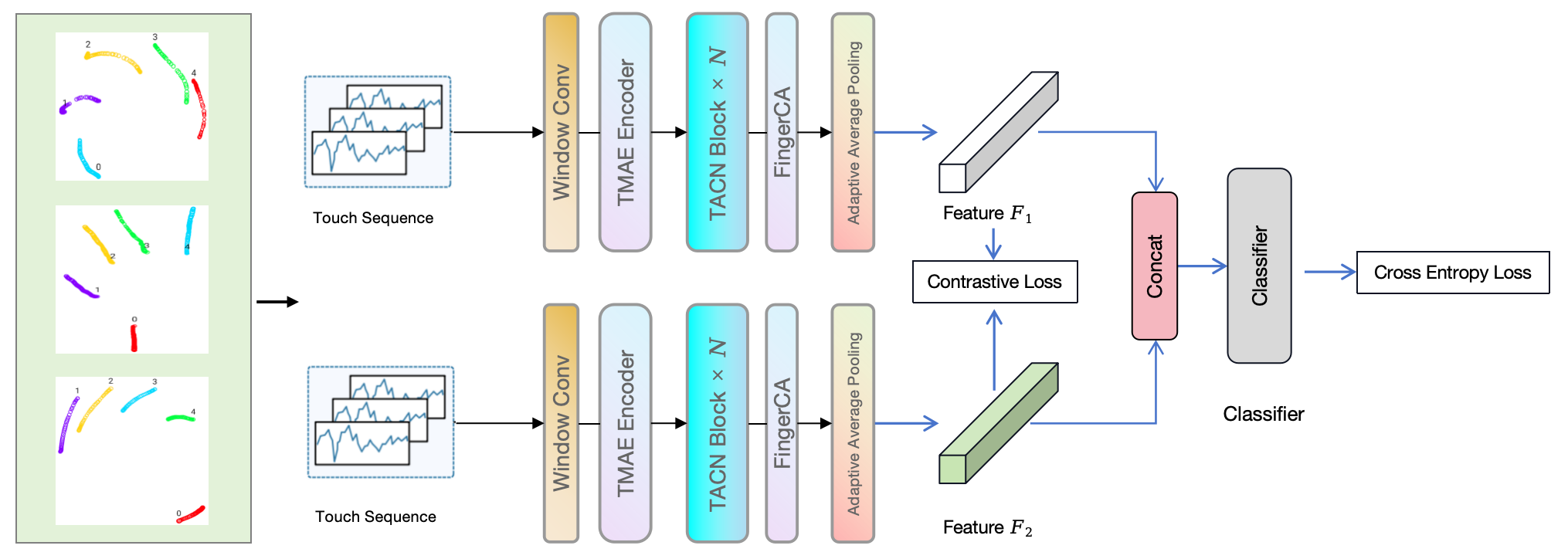}  
	\caption{TouchSeqNet}
        \Description{Mobile device user identity authentication model architecture TouchSeqNet}
	\label{fig:TouchSeqNet}
\end{figure*}

\subsubsection{Masked Representation Regression}

To reconstruct the representations of masked segments, we adopt a dual-encoder design consisting of a primary encoder and a momentum encoder. Both share the same Transformer architecture but differ in update strategies. 

The primary encoder encodes the visible input:
\begin{equation}
R_v = \text{Encoder}(Z_v)
\end{equation}

The momentum encoder, updated without gradients, encodes the masked tokens:
\begin{equation}
R_m = \text{Momentum\_Encoder}(Z_m)
\end{equation}

A cross-attention-based regressor predicts masked representations using visible context:
\begin{equation}
\hat{R}_m = \text{Regressor}(R_v, E_{\text{mask}})
\end{equation}

\subsubsection{Discrete Codeword Prediction}

To further enhance semantic learning, we introduce a discrete codeword prediction task. The tokenizer employs Gumbel-Softmax to discretize latent features. Given ground-truth discrete tokens \( T_m \in \mathbb{R}^{d_m} \), the model predicts codeword distributions from reconstructed embeddings:
\begin{equation}
P(\hat{T}_m) = \text{Tokenizer.center}(\hat{R}_m)
\end{equation}

\subsubsection{Momentum Encoder Updates}

The momentum encoder is updated via an exponential moving average of the primary encoder weights~\cite{chen2020simple}:
\begin{equation}
\theta_m \leftarrow \mu \theta_m + (1 - \mu)\theta_e
\end{equation}

Here, \( \theta_m \) and \( \theta_e \) denote the parameters of the momentum and primary encoders, respectively. A high smoothing factor \( \mu \) (e.g., 0.99) ensures stable target representations for regression.

\subsection{Self-supervised Loss}

The self-supervised objective of our model consists of two components: an alignment loss and a discrete codeword prediction loss.

The alignment loss \( \mathcal{L}_{\text{align}} \) minimizes the mean squared error (MSE) between the target representations from the momentum encoder and the predicted representations from the regressor:
\begin{equation}
\mathcal{L}_{\text{align}} = \text{MSE}(rep\_mask, rep\_mask\_prediction)
\end{equation}

The prediction loss \( \mathcal{L}_{\text{pred}} \) uses cross-entropy to measure the discrepancy between the predicted token distributions and the ground-truth discrete tokens. We also monitor auxiliary metrics such as \textit{Hits} and \textit{NDCG@10} for evaluation.

The final loss function is a weighted sum of the two:
\begin{equation}
\mathcal{L} = \alpha \mathcal{L}_{\text{align}} + \beta \mathcal{L}_{\text{pred}}
\end{equation}
where \( \alpha \) and \( \beta \) control the contribution of each term.

\section{TouchSeqNet Architecture}

As illustrated in Figure~\ref{fig:TouchSeqNet}, we propose TouchSeqNet, a contrastive learning framework designed for continuous user authentication based on dynamic touch data from mobile devices~\cite{lee2023soft}. The architecture integrates pretrained temporal encoder via self-supervised learning on unlabeled behavioral sequences, a Temporal-Attentive Convolutional Network (TACN) module, and a hybrid loss that combines contrastive and cross-entropy objectives.

By leveraging a contrastive learning paradigm, TouchSeqNet extracts user-consistent yet discriminative representations that generalize well to real-world usage conditions.

\subsection{Transfer Learning from TMAE}

TouchSeqNet leverages the strengths of the pretrained TMAE model by transferring its temporal encoder into the downstream identity authentication framework. Specifically, we reuse two key modules from TMAE: (1) a window-based convolutional projection layer; (2) a multi-layer Transformer encoder pretrained via self-supervised masked representation regression.

These two modules act as the feature extractor in TouchSeqNet, transforming raw touch sequences into latent representations that encode both local and contextual temporal patterns. This transfer learning strategy enables TouchSeqNet to initialize from a representation space aligned with touch dynamics and temporal structure. As a result, the model starts with a strong inductive bias tailored for behavioral biometrics, allowing subsequent layers to focus on refining identity-specific discriminative features.

\subsection{TACN block}

To capture both local and global temporal patterns in dynamic touch sequences, the TACN module combines Temporal Convolutional Networks (TCN)~\cite{hewage2020temporal} with Multi-head Attention to jointly model long-term dependencies and contextual correlations, while maintaining efficient and parallelizable computation. Compared with conventional CNNs, TCN supports longer effective memory via dilated convolutions without increasing parameter complexity~\cite{song2020distributed}.

\subsubsection{Temporal Convolutional Network}

The TCN is composed of stacked residual blocks, each containing two layers of dilated causal convolutions followed by weight normalization, ReLU activation, and dropout. A residual connection is added to stabilize training. The core computation in a residual block is given by:

\begin{equation}
	y_i = \text{ReLU}\left(W_2 * \left(\text{ReLU}(W_1 * X_i + b_1)\right) + b_2\right),
\end{equation}

where \( W_1, W_2 \in \mathbb{R}^{C_{\text{out}} \times C_{\text{in}} \times k} \) are convolutional kernels, and \( * \) denotes the dilated causal convolution.

Dilated convolutions allow exponential expansion of the receptive field. Given a kernel \( f \) of size \( k \) and dilation factor \( d \), the dilated convolution is defined as:

\begin{equation}
	F(s) = \sum_{i=0}^{k-1} f(i) \cdot X_{s - d \cdot i}.
\end{equation}

By setting \( d = 2^l \) for the \( l \)-th layer, the receptive field grows rapidly with depth while preserving the input length, which is essential for temporal alignment in authentication tasks.

Each residual block transforms the input sequence \( X \in \mathbb{R}^{C_{\text{in}} \times L} \) as:

\begin{equation}
	X^{(l)} = f(W^{(l)} *_{d} X^{(l-1)} + b^{(l)}), \quad X_{\text{out}}^{(l)} = X^{(l)} + X^{(l-1)}.
\end{equation}

This structure allows efficient learning of long-range patterns while maintaining temporal consistency across layers.

\subsubsection{Multi-head Attention and Hierarchical Temporal Fusion}

While TCN are well-suited for extracting local patterns, they may struggle to fully capture long-range temporal dependencies that span across the entire sequence. To address this, we incorporate a Multi-head Attention mechanism after the TCN layers to enhance the model's global temporal reasoning capabilities \cite{pan2022integration}. By leveraging multi-head attention, TACN can simultaneously attend to multiple temporal perspectives, enabling the model to better distinguish subtle variations in user touch dynamics, and laying a robust foundation for downstream identity authentication tasks.

\subsection{FingerCA}

\begin{figure}[t]
	\centering
	\includegraphics[width=0.5\textwidth]{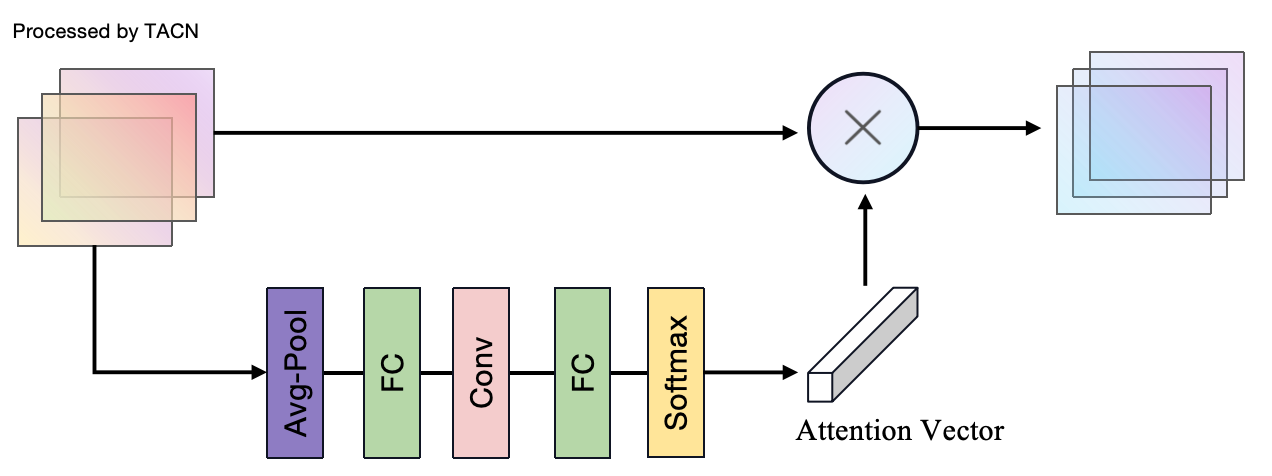}
	\caption{The structure of FingerCA module}
        \Description{FingerCA channel attention module}
	\label{fig:FingerCA}
\end{figure}

As illustrated in Figure~\ref{fig:Pre-training}, To enhance discriminability after temporal modeling, we incorporate the FingerCA channel attention module~\cite{hu2018squeeze, lin2024channel}. 

Let $X \in \mathbb{R}^{T \times C}$ be the output of the TACN blocks, where $T$ is the number of time steps and $C$ is the number of channels. A global average pooling is applied across time to produce a channel descriptor $z \in \mathbb{R}^{C}$:
\begin{equation}
z_c = \frac{1}{T} \sum_{t=1}^T X_{t,c}
\end{equation}
This vector is passed through a two-layer MLP with non-linear activation to obtain attention weights $\alpha \in \mathbb{R}^C$, which are used to recalibrate the input features: $\tilde{X} = \alpha \odot X$.

\subsection{Sample Pair Classification}
For each input pair, we obtain two representations $\tilde{F}_1$ and $\tilde{F}_2$, apply temporal average pooling, and concatenate the resulting vectors: $z = [z_1 \| z_2] \in \mathbb{R}^{2C}$. A fully connected classification head then predicts the probability that the pair belongs to the same user. This binary classification is trained end-to-end with a hybrid loss function.

\subsection{Hybrid Loss Function}

To jointly optimize representation learning and classification accuracy, we adopt a hybrid loss combining contrastive and cross-entropy terms~\cite{zahid2024multi}. Given embeddings $(z_1, z_2)$ with label $y \in \{0, 1\}$:

\begin{equation}
\mathcal{L}_{\text{contrastive}} = y \cdot \|z_1 - z_2\|^2 + (1 - y) \cdot \left[\max(0, m - \|z_1 - z_2\|)\right]^2
\end{equation}

\begin{equation}
\mathcal{L}_{\text{CE}} = -y \log(\hat{y}) - (1 - y) \log(1 - \hat{y})
\end{equation}

The final training objective is a weighted sum of both terms:

\begin{equation}
\mathcal{L}_{\text{total}} = \lambda_1 \cdot \mathcal{L}_{\text{contrastive}} + \lambda_2 \cdot \mathcal{L}_{\text{CE}}
\end{equation}

where \( \lambda_1 \) and \( \lambda_2 \) are hyperparameters balancing the two losses.

\section{EXPERIMENTS}

In this section, we describe our experimental protocol for evaluating TouchSeqNet. The evaluation consists of two stages: self‑supervised pretraining of the TMAE encoder on unlabeled touch dynamics data, and fine‑tuning the TouchSeqNet on labeled data. We conduct experiments on three datasets: our newly collected Ffinger dataset and two widely used public benchmarks, Touchalytics~\cite{frank2012touchalytics} and BioIdent~\cite{antal2015information}. 

\subsection{Experimental Setup}

\subsubsection{Data Description}

All datasets undergo the same preprocessing and normalization procedures to ensure consistency across experiments.

To handle variable-length sequences, we apply zero-padding along the temporal dimension. Specifically, for each sample $X \in \mathbb{R}^{T \times C}$, we pad it to length $T_{\text{pad}}$ such that $T_{\text{pad}} \bmod \sigma = 0$, where $\sigma$ is the convolution window size used in the pretraining stage. This ensures compatibility with window-based slicing.

We also construct a binary mask aligned with each sequence to mark valid positions. The mask is grouped into non-overlapping windows of size $\sigma$; a window is marked as masked if more than half of its positions are padding. If masking is required, this window-level mask is then used during TMAE pretraining to guide the selection of visible and masked windows.

For all datasets, we adopt a contrastive pairing strategy. Each input to the model consists of a pair of samples, with the following labeling scheme:
\begin{itemize}
  \item \textbf{Positive pair} ($y=1$): both samples belong to the same user.
  \item \textbf{Negative pair} ($y=0$): the two samples are drawn from different users.
\end{itemize}

This setup allows us to evaluate the model’s ability to learn identity-discriminative features under consistent experimental conditions.

\subsubsection{Evaluation Methods}

We assess the model using several standard classification metrics:

\begin{itemize}
	\item Accuracy: the proportion of correctly predicted sample pairs over all pairs.
	\item F1 Score: the harmonic mean of precision and recall, computed as
	\begin{equation}
	\text{F1} = \frac{2 \cdot P \cdot R}{P + R}.
	\end{equation}
	\item AUC: the Area Under the ROC Curve, indicating overall separability between positive and negative classes regardless of threshold.
\end{itemize}

\subsubsection{Parameter settings}

During the pretraining stage, we consistently set the embedding size to 64 for all models. The Adam optimizer is adopted as the default optimizer, with a fixed learning rate of 0.01 and no additional learning rate scheduling. The batch size is uniformly set to 128 across all experiments.

The Transformer encoder used in the TMAE model consists of 8 layers, each with 4 attention heads and a two-layer feed-forward network. A dropout rate of 0.2 is applied throughout the encoder \cite{eldele2021time}. In addition to the standard Transformer encoder, we further employ a 4-layer decoupled encoder to extract contextual representations from masked positions.

For each dataset, the slicing window size \(\delta\) is selected from the candidate set \(\{4, 8, 12\}\), and the default masking ratio is set to 40\%. The vocabulary size of the discrete codebook in the tokenizer is fixed to 192 in our implementation.

In the fine-tuning stage, all hyperparameters shared with the pretraining stage are preserved. Additionally, the Temporal Convolutional Network (TCN) used in TouchSeqNet is configured with the input dimension \texttt{num\_inputs} set to 64. The TCN consists of two residual blocks with output channel sizes defined as \texttt{num\_channels} = [64, 128], and a dropout rate of 0.2. For each dataset, the convolutional kernel size is selected from \(\{4, 5, 7\}\) based on validation performance.

\subsection{Experiment Results}

\subsubsection{Pre-training}

Recent advances in self-supervised learning~\cite{brown2020language} have shown great promise in learning transferable representations from unlabeled data~\cite{long2016deep}. In many domains, it is common to pre-train a model on a large dataset and fine-tune it on target tasks. However, in the context of touch dynamics, datasets often differ significantly in terms of acquisition methods, gesture types, device specifications, and behavioral protocols. 

To address this, we conduct self-supervised pretraining independently on the Touchalytics, BioIdent, and Ffinger datasets, and evaluate each model on its corresponding validation set. This setup ensures that the learned representations are adapted to the characteristics of each dataset and serve as a robust initialization for downstream fine-tuning.

\subsubsection{Evaluations on TouchSeqNet}

We evaluate the effectiveness of the proposed TouchSeqNet architecture on three representative touch dynamics datasets. Table~\ref{tab:classification_results} summarizes its performance on the held-out test sets. The results demonstrate that TouchSeqNet consistently achieves strong classification performance across different datasets, highlighting its robustness under diverse gesture and session conditions.

\begin{table}[htbp]
	\caption{Performance of TouchSeqNet on All Datasets}
	\label{tab:classification_results}
	\centering
	\begin{tabular}{lccc}
		\toprule
		\textbf{Metric} & \textbf{Ffinger} & \textbf{BioIdent} & \textbf{Touchalytics} \\
		\midrule
		Accuracy & 0.9769 & 0.9902 & 0.9908 \\
		F1 Score & 0.9770 & 0.9908 & 0.9907 \\
		AUC      & 0.9769 & 0.9907 & 0.9908 \\
		\bottomrule
	\end{tabular}
\end{table}

Across all evaluated datasets, TouchSeqNet consistently achieves high accuracy confirming its effectiveness in modeling fine-grained touch dynamics for identity authentication. On the public benchmarks Touchalytics and BioIdent, it delivers near-perfect classification performance, demonstrating strong robustness and generalization in cross-user scenarios.

These results highlight the model’s ability to extract user-specific and gesture-aware representations, supporting its deployment as a unified solution for continuous authentication in both controlled and real-world environments.

\subsubsection{Comparative Experiments}

To further evaluate the effectiveness of the proposed TouchSeqNet model, we conduct comparative experiments against several strong baseline models under a consistent contrastive learning framework. Each model is used as a feature extractor for paired inputs, and a downstream classification head makes binary decisions. All models are trained and evaluated under the same experimental settings, and their performance is measured using classification accuracy across three datasets: Ffinger (our dataset), BioIdent, and Touchalytics.

The baseline models include:

\begin{itemize}
	\item \textbf{TCN}: A dilated and causal convolutional network that captures long-range dependencies in time series data \cite{lea2017temporal}.
	\item \textbf{Gate-Transformer}: A lightweight attention-based model incorporating gating mechanisms to emphasize salient temporal features \cite{liu2021gated}.
	\item \textbf{LSTM}: A recurrent neural network that models sequential dynamics through memory cells and gating structures \cite{bajaber2022evaluation}.
	\item \textbf{InceptionTime}: A CNN-based model employing inception modules to extract multi-scale temporal features \cite{ismail2020inceptiontime}.
	\item \textbf{TSLANet}: A lightweight time series model featuring an Adaptive Spectral Block for Fourier-based denoising and an Interactive Convolution Block for efficient local feature extraction. \cite{eldele2024tslanet}.
\end{itemize} 

\begin{table*}[htbp]
	\centering
	\caption{Performance comparison of all models on Ffinger, BioIdent, and Touchalytics datasets (Accuracy / F1 / AUC).}
	\label{tab:comparison_all}
	\renewcommand{\arraystretch}{1.2}
	\begin{tabular}{lccc|ccc|ccc}
		\toprule
		\multirow{2}{*}{\textbf{Model}} &
		\multicolumn{3}{c|}{\textbf{Ffinger}} &
		\multicolumn{3}{c|}{\textbf{BioIdent}} &
		\multicolumn{3}{c}{\textbf{Touchalytics}} \\
		\cmidrule(lr){2-4} \cmidrule(lr){5-7} \cmidrule(lr){8-10}
		& Acc & F1 & AUC & Acc & F1 & AUC & Acc & F1 & AUC \\
		\midrule
		TCN\cite{lea2017temporal} & 0.9558 & 0.9582 & 0.9656 & 0.9052 & 0.9088 & 0.9050 & 0.9001 & 0.9033 & 0.9007 \\
		Gate-transformer\cite{liu2021gated} & 0.9343 & 0.9384 & 0.9332 & 0.9794 & 0.9789 & 0.9792 & 0.9852 & 0.9852 & 0.9852 \\
		LSTM\cite{bajaber2022evaluation} & 0.8851 & 0.8820 & 0.8812 & 0.9656 & 0.9638 & 0.9647 & 0.9713 & 0.9715 & 0.9713 \\
		InceptionTime\cite{ismail2020inceptiontime} & 0.8447 & 0.8644 & 0.8402 & 0.9886 & 0.9878 & 0.9885 & 0.9755 & 0.9754 & 0.9756 \\
		TSLANet\cite{eldele2024tslanet} & 0.7247 & 0.7453 & 0.7227 & 0.9831 & 0.9833 & 0.9831 & 0.9415 & 0.9426 & 0.9419 \\
		\textbf{TouchSeqNet (ours)} & \textbf{0.9769} & \textbf{0.9770} & \textbf{0.9769} & \textbf{0.9902} & \textbf{0.9908} & \textbf{0.9907} & \textbf{0.9908} & \textbf{0.9907} & \textbf{0.9908} \\
		\bottomrule
	\end{tabular}
\end{table*}

As shown in Table~\ref{tab:comparison_all}, the proposed \textit{TouchSeqNet} consistently outperforms all baseline methods across the three evaluated datasets. On the Ffinger dataset, it achieves the highest performance in all metrics, demonstrating strong robustness in modeling fine-grained and user-specific interaction patterns. In comparison, models such as TSLANet and LSTM show significant performance drops, indicating their limitations in capturing such behavioral variability.

On the public benchmarks BioIdent and Touchalytics, TouchSeqNet also reaches near-perfect results in terms of accuracy, F1 score, and AUC, outperforming or matching all competing approaches.

These results highlight the model’s superior generalization and discriminative capability, validating its effectiveness for continuous touch-based user authentication in practical scenarios.

\subsubsection{Ablation Study}

To assess the contributions of key components in TouchSeqNet, we conduct an ablation study across the Ffinger, BioIdent, and Touchalytics datasets. As summarized in Table~\ref{tab:ablation}, we compare the full model against three variants: (1) removing the multi-head attention in TACN (\textit{TACN w/o Attention}), (2) removing the Pretrained-module (\textit{TACN w/o Pretrained-module}), and (3) using only the Pretrained-module (\textit{Only Pretrained-module}).

Removing multi-head attention results in a clear drop in performance across all datasets (e.g., 97.69\% to 94.05\% on Ffinger), highlighting the importance of attention in capturing global temporal dependencies. Eliminating the Pretrained-module leads to an even more significant accuracy loss on BioIdent (from 99.02\% to 83.56\%) and Touchalytics, demonstrating the effectiveness of transfer learning from TMAE. Using only the Pretrained-module yields the lowest scores overall, confirming that pretraining alone is insufficient and must be complemented by downstream temporal modeling.

These results underscore the complementary benefits of the Pretrained-module and TACN components. While the encoder provides strong generalizable features, the attention-augmented temporal modeling in TACN is crucial for extracting task-specific discriminative representations.

\begin{table}[htbp]
    \small
    \caption{Ablation study results on classification accuracy}
    \label{tab:ablation}
    \centering
    \begin{tabular}{@{}lccc@{}}
        \toprule
        \textbf{Model Variant} & \textbf{Ffinger} & \textbf{BioIdent} & \textbf{Touchalytics} \\
        \midrule
        TACN w/o Attention & 0.9405 & 0.9806 & 0.9856 \\
        Only Pretrained-module & 0.9325 & 0.8733 & 0.9366 \\
        TACN w/o Pretrained-module & 0.9763 & 0.8356 & 0.8655 \\
        \textbf{TouchSeqNet (Full)} & \textbf{0.9769} & \textbf{0.9902} & \textbf{0.9908} \\
        \bottomrule
    \end{tabular}
\end{table}

\subsubsection{Summary of Experimental Findings}

Experimental results across Ffinger, BioIdent, and Touchalytics confirm the effectiveness and robustness of the proposed TouchSeqNet framework for dynamic touch-based authentication.

In comparative experiments, TouchSeqNet consistently outperforms strong baselines such as TCN, Gate-Transformer, LSTM, and InceptionTime. Its performance gains are most evident on the challenging Ffinger dataset, demonstrating its strength in modeling fine-grained temporal and identity-specific patterns.

Ablation results highlight the complementary roles of the core components. The multi-head attention mechanism in TACN enhances temporal discriminability, while the pre-trained module from TMAE provides transferable features that improve generalization, especially in low-data regimes. Removing either module leads to notable performance degradation.

Additionally, the model achieves near-perfect accuracy and F1 scores on public datasets, underscoring its strong generalization to different devices and behavioral contexts. The integration of contrastive learning and hierarchical temporal modeling enables robust discrimination between genuine and impostor pairs across a wide range of conditions.

\section{Conclusion}

We presented TouchSeqNet, a contrastive learning framework designed for continuous user authentication via touch dynamics. By integrating self-supervised pretraining, hierarchical temporal modeling, and attention mechanisms, the model learns rich, discriminative representations without requiring handcrafted features or domain-specific heuristics.

Our architecture demonstrates the synergistic value of transfer learning and structured temporal modeling. The pre-trained module offers a generalizable feature space, while the TACN block enhances temporal resolution and discriminability—together enabling robust identity verification across varying users and contexts.

Looking ahead, our framework provides a foundation for scalable behavioral biometrics. Its modular design makes it extensible to cross-device scenarios, federated authentication systems, and even multi-modal interaction signals (e.g., stylus, handwriting, or gesture input), pointing to broad applicability in real-world human-computer interaction systems.

\bibliographystyle{ACM-Reference-Format}
\bibliography{sample-base}
\end{document}